# Direct Growth Graphene on Cu Nanoparticles by Chemical Vapor Deposition as Surface-Enhanced Raman Scattering Substrate for Label-Free Detection of Adenosine

Shicai Xu[1*], Baoyuan Man[2], Shouzhen Jiang[2], Jihua Wang[1], Jie Wei[3], Shida Xu[4], Hanping Liu[1], Shoubao Gao[2], Huilan Liu[1], Zhenhua Li[1], Hongsheng Li[5], Hengwei Qiu[2]

[1] College of Physics and Electronic Information, Shandong Provincial Key Laboratory of Functional Macromolecular Biophysics, Institute of Biophysics, Dezhou University, Dezhou 253023, China

[2] College of Physics and Electronics, Shandong Normal University, Jinan 250014, China

[3] Department of Neurology, Dezhou People's Hospital, Dezhou 253014, China

[4] Department of Internal Medicine, Dezhou People's Hospital, Dezhou 253014, China

[5] Department of Radiation Oncology, Key Laboratory of Radiation Oncology of Shandong Province, Shandong Cancer Hospital and Institute, Jinan 250117, China

[*] E-mail address: xushicai001@163.com (S. C. Xu)

**Abstract**

We present a graphene/Cu nanoparticle hybrids (G/CuNPs) system as a surface-enhanced Raman scattering (SERS) substrate for adenosine detection. The Cu nanoparticles wrapped around a monolayer graphene shell were directly synthesized on flat quartz by chemical vapor deposition in a mixture of methane and hydrogen. The G/CuNPs showed an excellent SERS enhancement activity for adenosine. The minimum detected concentration of the adenosine in serum was demonstrated as low as 5 nM, and the calibration curve showed a good linear response from 5 to 500 nM. The capability of SERS detection of adenosine in real normal human urine samples based on G/CuNPs was also investigated and the characteristic peaks of adenosine were still recognizable. The reproducible and the ultrasensitive enhanced Raman signals could be due to the presence of an ultrathin graphene layer. The graphene shell was able to enrich and fix the adenosine molecules, which could also efficiently maintain chemical and optical stability of G/CuNPs. Based on G/CuNPs system, the ultrasensitive SERS detection of adenosine in varied matrices was expected for the practical applications of medicine and biotechnology.

**Keywords:** Graphene, Cu nanoparticles, SERS, Adenosine detection

# 1. Introduction

In recent decades, surface-enhanced Raman spectroscopy as a very important analytical technique for biomedical detection has received increasing attention.[1-3] As a widely used method, SERS shows its unique and excellent properties for the biological system. Although Raman spectroscopy is limited by low sensitivity, SERS could provide signal intensity of the molecules enhanced by orders of magnitudes on the proper substrates. During the past decades, noble metallic nanomaterials, such as Ag, Au and Cu nanoparticles, as SERS-active substrates have been widely reported.[4] Numerous attempts have been done to design well-ordered Ag or Au nanostructures with high SERS activity.[4,5] Despite considerable efforts, it is still a challenge to achieve ideal SERS substrates with good stability and reproducibility.[6] Actually, the issue of metal–molecule contact induced signal variations has become a key problem for practical applications.[7,8] Furthermore, the lower adsorption capacity of metal nanostructures for some molecules often limits their applications.[9]

Using a thin and pinhole free layer of $SiO_2$ or $Al_2O_3$ as an inert shell to isolate metal nanostructures from their surroundings was demonstrated,[7] but the approach are challenging for a normal metal substrate. Fabrication of SERS substrates with an ultrathin passivated surface at a lowest loss of electromagnetic enhancement activity is the key to shell-isolated SERS. The main challenge is to get a pinhole-free coating layer with a very small thickness. The atomic thickness and seamless structure of graphene makes it a natural candidate material for shell-isolated SERS. Graphene, a 2D atomic crystal with densely packed carbon atoms in a honeycomb crystal lattice,[10] is well-known for its unique electrical performance and the amazing applications in nanoscale electronics.[11] As well as the wide interest in its electric properties, graphene is also a rising star in Raman spectroscopy[12]. An atomic flat surface of graphene causes a small-distance charge transfer between the graphene surface and adsorbed molecules, making the Raman signal more reliably and efficiently. Beside the ultrathin shtructure, graphene also has large specific surface area of 2630 $m^2/g$,[13] which could act as an excellent adsorbent towards organic molecules, especially the aromatic molecules. Therefore, graphene can also work as a molecule enricher in SERS-active substrate.[14,15] It has been reported that graphene can effectively enhance the Raman signal and reduce the back action noise.[16,17]

Recently, many studies have been done to obtain graphene–noble metallic nanomaterials for SERS. These hybrids show great promise for applications in SERS. Ren et al. reported an sensitive SERS substrate for folic acid detection using graphene oxide/Ag nanoparticle hybrids.[18] He et al. demonstrated that the gold decorated graphene can serve as a SERS-active substrate for multiplexing detection of DNA.[19] Murphy et al. reported an enhanced sensitivity for SERS detection based on reduced graphene oxide (rGO)–Ag nanoparticle hybrid.[20] Hu et al. assembled SERS platform using graphene oxide (GO) and gold nanorod hybrids though electrostatic self-assembly procedure.[21] Qian et al. reported silver nanoparticles and GO sheets hybrid composite with good SERS performance.[22] Tang and co-workers demonstrated a distinct SERS effect based on the silver–GO or rGO materials fabricated by using glucose as the reducing agent.[23] Dutta et al. proposed a wet-chemical method to prepare Ag nanoparticle conjugated rGO nanosheets for uranylion detection.[24] These works provided facile methods to decorate metal nanoparticles on the surface of graphene (or GO) on which the graphene can be as an effective molecule enricher. However, the adopted graphene layer in there methods did not essentially isolate the probed materials from the metallic Raman reporters. To solve this problem, Xu et al. prepared a graphene-veiled gold substrate with a passivated surface for SERS.[25] Xie's group fabricated a hybrid SERS-active platform consisting of a single layer graphene (SLG) covering a quasiperiodic Au nano-pyramid and showed a very high SERS enhancement factor of over $10^{10}$ for rhodamine 6G (R6G).[26] Zhang's group has shown that few-layer graphene-encapsulated metal nanoparticles hybrid is a promising material for shell isolated SERS.[27] Very recently, Chen et al developed a facile method to fabricate GO encapsulated Ag particles hybrid material as SERS probe.[28] More reproducible SERS signals were demonstrated in these studies by employing graphene or GO nanosheets as the passivated surface. Nevertheless, there is still a principal disadvantage in signal stability and reproducibility for these methods. As in all the above mentioned methods, the graphene-metal hybrids were obtained either through transferring a graphene sheet on the surface of metal nanoparticles or spin-coating a mixed solution of graphene and metal nanoparticles. As both approaches essentially belong to physical composition, the tightly sealed structure is hard to form between metallic nanoparticals and graphene shell. The space between graphene and metal nanoparticles will inevitably cause apparent loss of electromagnetic enhancement activity as the electromagnetic enhancement efficiency decays rapidly with distance between the metal nanoparticles and analytes. Moreover, the suspended and wrinkled graphene structures always formed nearby the gaps of nanoparticals,[26,28] making graphene-metal hybrids non-uniform and easy to damage, further reducing in the homogeneity and reproducibility of the SERS substrates. It is satisfactory to use chemical vapor deposition (CVD) method to directly grow a thin layer of graphene on the surface of nanoparticals. High quality graphene can be grown on Cu foil at the high temperature of ~1000 ℃.[29] However, for metal nanoparticals (Au, Ag and Cu), due to the small size effect they can not withstand the high graphene growth temperature, at which these nanoparticals will almost be vaporized or at least completely melted depending on their original shape and size. Thus, it is desired to develop a new way to grow graphene layer on metal nanoparticals to fabricate metal–graphene SERS substrates with good stability and reproducibility.

In this work, we provide a direct growth approach to prepare a high performance SERS substrate with graphene-wrapped Cu nanoparticals by a two-temperature-zone CVD method. As a chemistry composite mode, this method provides an atomically thin, seamless, and chemically inert net to tightly wrap the metallic nanoparticals. Compared with previously reported works, this approach exhibits three advantages. Firstly, the direct growth mode make graphene layer tightly wrap the metal nanoparticals, minimize the loss of electromagnetic enhancement activity. Secondly, the direct grown graphene shell is free from suspended and wrinkled structures and effectively avoids graphene damage, thus providing enhanced chemical stability and reproducibility of SERS. Thirdly, direct grown graphene can cover every place of the G/CuNPs, even the narrow gaps of particles (hot spots). So the target molecules can be more effectively absorbed on the hot spots and thus more sensitive Raman signals are expected. Based on the G/CuNPs, ultrasensitive and label-free detection of adenosine by SERS spectra was demonstrated. Adenosine, a metabolite of adenine nucleotides, is one of the major neuromodulators. Adenosine has been recognized as an endogenous anticonvulsant and neuroprotective molecule. As the core molecule of ATP and of nucleic acids, adenosine forms a unique link among cell energy, gene regulation, and neuronal excitability.[30] Adenosine has long been a highly coveted therapeutic target, and its actions at the A1 receptor subtype hold well-established and profound therapeutic potential for conditions such as stroke, brain injury, pain, and epilepsy, among others.[31,32] Adenosine also has drawn attention as possible tumor markers in human cancers, because adenosine in cancer patients is generally found to be significantly higher than that in healthy person.[33-35] Therefore, it is interesting to develop a highly specific and sensitive detection method for the practical applications of medicine and biotechnology. On the basis of the G/CuNPs with strong SERS enhancement and enrichment of adenosine, low limit of detection was achieved in water, serum and normal human urine by SERS. We succeeded in collecting clean and more reproducible SERS signals in varied matrices, displaying great potential for the practical applications of medicine and biotechnology.

**RESULTS AND DISCUSSION**

The prime feature of the proposed approach include fabrication of Cu nanoparticles (CuNPs) by remote thermal evaporation and the direct formation of graphene layer on the surface of CuNPs in the mixture of Cu vapor, H and decomposed $CH_4$. Figure 1a shows the growth setup of G/CuNPs. The CVD hot-wall reactor was designed with two temperature zones in which the temperature can be controlled independently. For G/CuNPs growth, the reactor was set with one zone at 600 ℃ and the other at 1050 ℃. A strip of Cu foil cleaned by acetic acid surrounding along the tube wall was placed in the high-temperature zone in the upstream gas flow to supply CuNPs and achieve graphene growth in a mixture of $H_2$ and $CH_4$. The quartz substrates were placed in the low-temperature zone downstream at ~10 cm away from the Cu foils for G/CuNPs deposition. Figure 1b schematically illustrates the working mechanism of the graphene growth on the surface of CuNPs. The Cu foil sublimes and produce a large number of Cu atoms at the high temperature of 1050℃ at low pressure of 280 Pa. When the Cu atoms attain a certain concentration, these Cu atoms begin to merge with each other, evolving into Cu nanoparticles with a certain size. On the other hand, the Cu atoms are also used as catalysts to decomposes $CH_4$, enabling a typically CVD reaction to grow a graphene layer on floating CuNPs. The graphene-wrapped Cu nanoparticles are then transported downstream with gas flow into the low-temperature zone, forming G/CuNPs structure on quartz substrates. As a comparison, we also prepared single CuNPs in the mixture of $H_2$ and Ar. The traditional CVD graphere grown on Cu foil were also transferred on the Cu nanoparticles, making the transferred graphere/Cu nanoparticle hybrids (TG/CuNPs). The detailed description of the experiment process can be found in methods.

Scanning electron microscope (SEM) was used to investigate the surface morphology of G/CuNPs formed on the quartz substrate. Figure 2a shows the image of sample at a low magnification. The quartz substrate is covered by a large number of nanoparticles with uniform size. To identify the size of nanoparticles, a higher magnification was adopted. As shown in Figure 2b, the average size of these nanoparticles is ~100 nm and the gaps among nanoparticles are very narrow. Since metallic nanoparticles with narrow gaps are regarded to be essential for large Raman enhancement,[36] the narrow gaps are expected to support huge electromagnetic enhancement for absorbed molecules. As the direct grown graphene layer can not be directly recognized by SEM observations, a conventional Raman measurement was performed on these nanoparticles to demonstrate the existence of graphene. As shown in Figure 2c, the D, G and 2D peaks of graphene are clearly observed at ~1360, ~1580 and ~2695 cm$^{-1}$, respectively. These peaks can be regarded as the fingerprint of graphene.[37,38] The Raman spectrum shows typical features of monolayer graphene: the intensity ratio of $I_{(2D)}/I_{(G)} \geq 2$ and a single Lorentzian 2D peak with a full width at half maximum of 30~40 cm$^{-1}$.[29,39] Here, the defect-related D peak is very weak, indicative of high quality of graphene. These results demonstrate that the graphene film with a monolayer structure is actually grown on nanoparticals. By replacing the mixture of H$_2$ and CH$_4$ by H$_2$ and Ar, the single CuNPs are also formed on the quartz substrates (Figure 2d). The CuNPs have similar size and distribution as G/CuNPs. As a control experiment, we also fabricated TG/CuNPs by a transferring a piece of CVD Cu foil based-graphene on the CuNPs. As shown in Figure 2e, the boundaries of nanoparticles become blurred after adding the graphene layer. In fact, because of the large fluctuation of the nanoparticles, the transferred graphere film had difficulty in getting closer to the surface of nanapaticals.[40] As shown in the inset of Figure 2e, the suspended and wrinkled graphene are formed on the edge of the nanoparticles. These suspended and wrinkled graphene structures were also appeared in the SEM image of graphere/nanapaticals hybrids in previous studies.[26,28] It is anticipated that these suspended and wrinkled graphene structure make the electromagnetic "hot" spots large distance from adsorbates and will cause apparent loss of electromagnetic enhancement active. On the other hand, the suspended structure can be damaged more easily, further resulting in the inhomogeneity of SERS substrates. Figure 2f shows the Raman spectrum collected from the TG/CuNPs. The typical graphene peaks are also observed. However, compared to Raman spectrum of the as-grown graphene in the G/CuNPs, the defect-related D peak is much higher. The high level of defect can be attributed to the transfer damage in the graphene transfer process.

The high resolution transmission electron microscopy (HRTEM) was employed to identify the exact dimensions of the G/CuNPs. Figure 3a shows the bright-field TEM image of G/CuNPs. The as-prepared nanoparticals are spherical shape with average diameter of ~80 nm. Considering electromagnetic enhancement efficiency decays rapidly with distance R between donor and acceptor under a $1/R^{12}$ or $1/R^{10}$.[41]

Ultrathin shell is a key factor for generating high SERS activity. In order identified the thickness of graphene shell, very detailed structural information nearby interface was captured by cross section HRTEM image of G/CuNPs (Figure 3b). It is apparent that Cu particles are covered by an ultrathin film, suggesting the formation of graphene-wrapped Cu nanopartical hybrids. The shell layer is about 0.34 nm in thickness, indicative of monolayer graphene.[10] The graphene layer can be further characterized by removing the Cu nanoparticals in 0.5 M $FeCl_3$ solution. As shown in Figure 3c, lance-shaped graphene sheets are observed in bright-field TEM of the graphene sheet after removing the Cu nanoparticals. This result can be understood by considering that the spherical-graphene shell can not be self-supported after removing the inner Cu nanopartical. The average size of lance-shaped graphene sheet is ~100 nm, slightly lager that that of the G/CuNPs. This phenomenon can be attributed to the geometry change of graphene shell. A HRTEM analysis of folds at the edges of lance-shaped graphene sheet can give the number of graphene layers by direct visualization. The number of dark lines represents the number of graphene layer.[42] As shown in Figure 3d, the HRTEM image derived from the edge of the lance-shaped graphene sheet exhibits only one dark line, indicative of monolayer graphene. Figure 3e shows the SAED patterns of graphene sheet. Typical six-fold symmetry patterns are observed from the region marked with the white circle in Figure 3c, indicating the single-crystalline nature of the observed domain.[43] For further quantitative analysis of diffraction patterns, we labeled the peaks with Bravais-Miller indices. As shown in Figure 3f, the inner peaks ($0\bar{1}10$) and ($\bar{1}010$) are more intense than the outer ones

($1\bar{2}10$) and ($\bar{2}110$), further confirming monolayer nature of the graphene.[44]

XPS is a kind of surface sensitive technique to analyze the chemical composition. Figure 4a shows the full XPS spectrum of G/CuNPs on quartz substrates. The characteristic peaks from Cu high are clearly observed at ~935 and ~952 eV, indicating the formation of Cu nanoparticles.[45] The signals of O1s at ~532 eV, Si 2s at and Si 2p at ~105.1 eV are assigned to quartz substrate.[45,46] The distinct peak of C1s at ~ 284 eV is the signature of $sp^2$ C–C network, consistent with the formation of graphene.[47] Figure 4b shows the detailed C1s core-level spectra of G/CuNPs. The C1s can be deconvolved into three components. The main peak at 284.5 eV indicates the formation of a $sp^2$ C–C network for the grown film.[48] The discernible tail at 286.3 and 288.9 eV is assigned to the hydroxyl carbon C–O and C=O, respectively.[47] The higher energy C1s peaks related to carbon-oxygen bonds are often observed in the CVD grown graphene, which is probably due to the oxygen contamination inside of the growth chamber. In a further comparison with transferred gaphene on $SiO_2$,[49] the C1s peak exhibits a slightly blueshift (from 284.4 eV to 284.5 eV). We suppose that weak chemical bonding interaction between graphene and CuNPs causes descreening of nucleus charges, leading to an overall increase in core electron binding energies.

To estimate the SERS activity of the G/CuNPs, $10^{-6}$ M aqueous solution of adenosine was chosen as the probe molecule. The SERS spectra of adenosine on the TG/CuNPs and CuNPs were also collected as the contrast. As shown in Figure 5a, three sets of bands are observed on the SERS spectra of adenosine on three kinds of SERS substrates. The primary vibrations of adenosine are confirmed according to the reported work.[50-53] It is distinct that the intensities of SERS spectra on G/CuNPs are much stronger than those of TG/CuNPs or CuNPs. The peaks at 725, 1483, 1508 and 1576 cm$^{-1}$ assigned to the ring breathing modes of the whole molecular from G/CuNPs is about 2-5 times stronger than that of TG/CuNPs and 15-20 times stronger than that of CuNPs. The peak at 847 cm$^{-1}$ assigned to skeletal mode of C-O-C from G/CuNPs is ~2.5 times stronger than that from TG/CuNPs and ~17.5 times stronger than that from CuNPs. The peak at 1307 cm$^{-1}$ assigned to the stretching vibration of N–C–N and C–C–N from G/CuNPs is ~2.2 times stronger than that from TG/CuNPs and ~11.3 times stronger than that from CuNPs. The peak at 1249 and 1337 cm$^{-1}$ assigned to the stretching vibration of C–N and the bending vibration of C–H from G/CuNPs is ~2.2 times stronger than that from TG/CuNPs and ~13.6 times stronger than that from CuNPs. The peak at 1371 cm$^{-1}$ assigned to bending vibration of C–H, N–H and the stretching vibration of C-N from G/CuNPs is ~2.0 times stronger than that from TG/CuNPs and ~12.1 times stronger than that from CuNPs. The peak at 1390 cm$^{-1}$ assigned to CH rocking from G/CuNPs is ~1.4 times stronger than that from TG/CuNPs and ~8.8 times stronger than that from CuNPs. Obviously, the sharp characteristic peaks in SERS spectrum of adenosine from G/CuNPs exhibits the best signal-to-noise ratio. Compared to the CuNPs, the graphene-wrapped CuNPs, both G/CuNPs and TG/CuNPs have better SERS activity. The additional enhancement of SERS signal of adenosine on graphene-wrapped CuNPs can be attributed to the molecule enrichment from graphene and graphene-derived CM enhancement. In a further comparison with TG/CuNPs, the better enhancement from G/CuNPs can be attributed to the tight combination between graphene and Cu nanoparticals. Because of direct growth mode, the graphene is free from suspended and wrinkled structure and have a more close contact with the inside nanoparticals. Since the electromagnetic enhancement decays rapidly with distance between metal and analyte,[41] the close contact structure makes target molecules closer to the surface and thus is more suitable for large SERS activity. On the other hand, the growth combination mode avoids the damages induced by the transfer process, the chemical enhancement mechanism from graphene associated with charge transfer effects are expected to be enhanced. Furthermore, we tested stability of the SERS substrate based on G/CuNPs, TG/CuNPs and CuNPs though a thermal oxidation treatment. The thermal oxidation treatment was implemented by expositing G/CuNPs, TG/CuNPs and CuNPs to hot and humid air (temperature: 85 °C, humidity: 80%) for 240 h. As shown in Figure 5b, after the oxidation treatment, the SERS intensity of G/CuNPs the adenosine is nearly unchanged, indicating excellent chemical stability of G/CuNPs. For the TG/CuNPs, the SERS intensity of the adenosine is obviously decreased after oxidation treatment. While for the oxidation treated CuNPs, the SERS intensity of the adenosine become very weak and some of the characteristic peaks of adenosine are absent. It has been known that CuNPs are easily oxidized when it is exposed to air. The decrease of the Raman signals can be ascribes to the formation of copper oxides on the surface of Cu nanoparticals. This assumption was confirmed by X-ray diffraction (XRD), energy dispersive spectroscopy (EDS) measurements and X-ray photoelectron spectroscopy analysis of the samples (Figure S1 and S2, Supporting

Information) after the thermal oxidation stability test. As shown in Figure S1a, the typical peaks of $Cu_2O$ (111) and CuO (100) are observed for G/CuNPs.[54] For the TG/CuNPs, weak peaks of $Cu_2O$ (111) and CuO (100) are also observed.[54] While for the G/CuNPs, the signal of copper oxide is negligible. The great change in the morphology of the CuNPs after the thermal oxidation indicates serious oxidative damage (Figure S1b), which corresponds to obvious oxygen increase in the EDS of G/CuNPs (Figure S1c). XPS measurements of the samples also shows shake-up features at~945 and ~965 eV for the Cu $2p_{3/2}$ and $2p_{1/2}$ core levels, which are evident and diagnostic of an open $3d^9$ shell of Cu (+2) (Figure S2, Supporting Information).[55] The formation of copper oxide on the surface of CuNPs increases the thickness of the passivation layer, which results in decreasing enhancement activity of samples. As oxygen gas and moisture cannot permeate through the graphene layer, the graphene can effectively protect metal from oxygen damage. The thermal stability comparison indicates that the grown graphene shell on CuNPs can more effectively suppress degradation of the metallic nano-structures in comparison with the transferred graphene shell. The grown graphene shell on CuNPs avoids suspended structure and transfer-induced damages on graphene layer, exhibiting better antioxidant ability for SERS substrates.

To achieve a lower limit of detection, the strongest peaks located at 847 and 1337 $cm^{-1}$ were chosen as the signature to determine the concentration of adenosine in the samples. SERS spectra of the adenosine in a series of concentrations tested on G/CuNPs, TG/CuNPs and the CuNPs substrate are shown in Figure 6a, b and c, respectively. The intensities of the SERS spectra rise with the increasing concentrations of adenosine, suggesting that the intensities are proportional to the amount of adenosine molecules. The minimum concentration of adenosine detected from G/CuNPs was as low as $10^{-9}$ M, which is one order of magnitude lower than that from TG/CuNPs and two orders of magnitude lower than that from CuNPs. The adenosine detection limit based on G/CuNPs is also much lower than that of other reported SERS adenosine detection or other methods for adenosine detection.[56-59] These results indicate that ultrasensitive detection of adenosine can be achieved based on the G/CuNPs. The SERS intensity of the vibration located at 847 and 1337 cm$^{-1}$ versus the concentration of adenosine are also plotted in Figure 6d. Possibly because of the non-uniformity of G/CuNPs, the standard deviations of the intensity for some concentrations are relatively large. Nonetheless, a good linear SERS response from $10^{-9}$ to $10^{-4}$ M of adenosine is obtained. The coefficient of determination ($R^2$) of the linear fit calibration curve for the peaks of 847 and 1337 cm$^{-1}$ is reached 0.995 and 0.998, respectively. The linear SERS responses versus concentration of adenosine were also obtained on TG/CuNPs and CuNPs from $10^{-8}$ to $10^{-4}$ M and from $10^{-7}$ to $10^{-4}$, respectively (Figure S3, Supporting Information). For both cases, the linear correlation between SERS spectra intensity and adenosine concentration are not as good as that collected on G/CuNPs (much smaller $R^2$ than that from G/CuNPs shown in Figure S3). It is indicated that G/CuNPs can provide more reliability and stability SERS signals for adenosine detection. The improved stability of SERS signals can also be attributed to the additional grown graphene layer. Compared to the transferred graphene, the as-grown graphene are free from suspended structure and transfer-induced damage, providing more uniform and effective adsorption site for adenosine.

To investigate the feasibility of the detection of adenosine in real biological samples, the adenosine from 5 to 500 nM was added to the diluted human serum (one percent of serum in water). Then the diluted serum containing adenosine was detected on G/CuNPs. The SERS spectra of adenosine in different concentrations in diluted serum are illustrated in Figure 7a. The intensities of the SERS spectra of adenosine are proportional to the concentration of the adenosine in diluted serum. Besides, the characteristic SERS spectra of adenosine in serum are quite similar to that in water with comparable intensity. There are only a few weak additional peaks in the spectrum derived from blank serum (black curve in Figure 7a). These weak additional peaks are attributed to the components in serum which do not disturb the recognition of adenosine. It can be deduced that the influence of the remaining protein in serum is almost ignorable in adenosine detection. The lowest detected concentration of adenosine in serum is 5 nM, corresponding to the spectrum (red curve in Figure 7a). To represent the capability of the quantitative detection of adenosine in serum and its reproducibility, the linear fit calibration curve ($R^2$ = 0.993) with error bars is illustrated in Figure 7b. The reasonable linear response of SERS is observed from 5 to 500 nM. The concentration gradient experiments of adenosine proved that the obtained G/CuNPs are good SERS substrates for the detection of adenosine in serum and the ignorable protein background indicates a potential application to detect adenosine in other practical biological systems.

As adenosine is also a possible tumor marker in human urine, we tested the normal human urine by using G/CuNPs. Figure 8a shows the SERS spectrum collected from human urine. The strong peaks at ~1002 cm$^{-1}$ is assigned to symmetric ring breathing mode of phenylalanine.[60,61] The strong peak at ~1436 cm$^{-1}$ is assigned to CH$_2$ scissoring of lipids and the one at ~1460 cm$^{-1}$ is assigned to CH$_2$/CH$_3$ deformation lipids and collagen.[60,62] The strong peaks at ~1593 cm$^{-1}$ is assigned to C=N, C=C stretching and ring stretches of Phenylalanine.[63] The distinct peak at ~1153 cm$^{-1}$ is assigned to carbohydrates for solutions.[53] It is worth noting that the adenosine related peaks at ~725, ~847, ~1307, ~1337, ~1371, ~1390, ~1483 and 1508 cm$^{-1}$ can also be clearly recognized in the real human urine. The ~725 cm$^{-1}$ assigned to ring breathing mode of adenosine is much enhanced, which can be attributed to the contribution of creatinine.[53] More clear Raman peaks of adenosine can be obtained by extracting nucleosides from the urine samples. As shown in Figure 8b, the adenosine related peaks are much enhanced. As the adenosine excreted in the urine of patients with malignant tumours is usually in abnormal levels. The SERS spectra of adenosine obtained from G/CuNPs shows great potential in incipient cancer diagnosis and in the monitoring of therapeutic effects. However, the G/CuNPs based SERS substrate is still need to be improved for practical applications. It is expected to obtain better reproducibility signal by improving the uniformity of nanoparticles by tuning the growth procedure of G/CuNPs. As the metallic silver has better plasma characteristics than copper, more sensitive SERS substrate is expected via replacing inner Cu nanoparticals by Ag nanoparticals. In fact, direct growth of monolayer graphene on metallic silver substrate is also possible by using solid carbon source according to the recent reports.[64] Further studies are now in progress in our group.

**CONCLUSION**

Herein graphene-wrapped Cu nanopartical hybrids (G/CuNPs) were prepared as SERS substrates according to direct growth method. On the basis of the obtained G/CuNPs, an ultrasensitive and label-free SERS strategy was developed for the detection of adenosine in water, serum and normal human urine according to the inherent SERS spectra of adenosine. The graphene shell was used to enrich and fix the adenosine molecules, on which reproducible and the ultrasensitive SERS signal of adenosine was obtained. The contrast of SERS spectra of adenosine on G/CuNPs, TG/CuNPs and CuNPs showed that the direct graphene growth mode on CuNPs was very important for the ultrasensitivity and reproducibility of the SERS detection of adenosine. It was demonstrated that the minimum detected concentration of the adenosine in serum was as low as 5 nM, and the calibration curve showed a good linear relation with a linear response from 5 to 500 nM. The capability of SERS detection of adenosine in real normal human urine samples based on G/CuNPs was also investigated and the characteristic peaks of adenosine were still recognizable. The versatility of this ultrasensitive SERS detection of adenosine in varied matrices was expected for the practical applications of medicine and biotechnology.

**EXPERIMENTAL SECTION**

**Fabrication of G/CuNPs on flat quartz substrate.**

For G/CuNPs fabrication, the CVD hot-wall reactor was set with the low-temperature zone at 600 °C and the high-temperature zone at 1050 °C. A strip of Cu foil cleaned by acetic acid was placed surrounding along the tube wall in the high-temperature zone in the upstream gas flow to supply CuNPs. The graphene films were grown on the floating CuNPs by CVD process in a mixture of $H_2$ and $CH_4$. The quartz substrates were placed in the low-temperature zone downstream at ~10 cm away from the Cu foils for G/CuNPs deposition. After the vacuum reached a pressure of $10^{-5}$ Pa, the tube was rapidly heated up to 1050 °C with a rate of ~100 °C/min with flowing 50 sccm $H_2$ and 150 sccm Ar at 1350 Pa. The mixture of $H_2$ and $CH_4$ was used to remove the oxide layer on Cu foil and restrain the Cu evaporation during heating process. When the high-temperature zone reached 1050 °C, a mixture of 50 sccm $H_2$ and 10 sccm $CH_4$ was introduced into the tube to replace the mixture of $H_2$ and Ar at a low pressure of 280 Pa to start graphene growth. After 5 min CVD reaction, the G/CuNPs were deposited onto the flat quartz substrate in the low-temperature zone. Finally, the $CH_4$ was shut off and the quartz tube was rapidly cooled down to room temperature with flowing 50 sccm $H_2$ and 150 sccm Ar at 1350 Pa. The detailed experimental parameters were illustrated in Figure S4 in the Supporting Information. Unlike the self-limited growth mode on Cu foil, muti-layer graphene was commonly grown on CuNPs. In order to obtain monolayer graphene on CuNPs, the $H_2/CH_4$ ratio was tuned. The monolayer graphene with low level of defect was achieved with 50 sccm $H_2$ and 10 sccm CH$_4$ (Figure S5, Supporting Information).

**Fabrication of CuNPs on flat quartz substrate.**

For CuNPs fabrication, the whole procedure was similar to that of G/CuNPs fabrication. The difference here is that when the high-temperature zone reached 1050 °C the mixture of 50 sccm H$_2$ and 150 sccm Ar at 1350 Pa was replaced by 50 sccm H$_2$ and 10 sccm Ar at 280 Pa (Figure S6, Supporting Information). After 5 min reaction, the CuNPs were deposited onto the flat quartz substrate in the low-temperature zone.

**Fabrication of TG/CuNPs.**

The CVD monolayer graphene was grown on 25 μm Cu foil at 1050°C with flowing of 60 and 15 sccm by using the growth procedure in our recent report.[65] A 200 nm-thick PMMA was deposited onto the graphene film by the spin-coated method and then the Cu foil was etched away by 0.5 M aqueous FeCl$_3$ solution. After removing the residual etchant in deionized water, the PMMA-coated graphene films were transferred onto the CuNPs layer. In order to relax the underlying graphene, a second PMMA coating was introduced onto the precoated PMMA layer. Then, the whole PMMA layer was dissolved with acetone, forming the TG/CuNPs structure. Finally, the TG/CuNPs were dried at 50 °C for 30 min with flowing 100 sccm Ar to evaporate the solvent completely.

**Preparation of urine samples.**

The single, early-morning urine samples were provided by Department of Internal Medicine of Dezhou People's Hospital from ten healthy volunteers (physical examination people). The samples were centrifuged at a speed of 10000 rpm for 10 min and the upper liquid was collected for Raman measurements. The nucleosides were extracted from urines by affinity chromatography using a phenylboronic acid gel (Affi-gel 601) in a glass column. After the gel was activated and equilibrated with 20 ml of 0.1 M NH$_4$OAc, 1 ml centrifuged urine was applied to the column. Then the gel was washed with 10 ml 0.1 M NH$_4$OAc and 10 ml methanol-water (1:1, v/v). The nucleosides were eluted with 10 ml 0.05 M HCOOH in methanol-water (1:1, v/v). Finally, the nucleoside solution was enriched to 1 ml by evaporation for Raman measurements.

**SERS Experiments.**

SERS experiments were carried out using a Raman system (Horiba HR-800) with laser excitation at 532 nm (2.33 eV). The excitation laser spot was about 0.5 μm and the incident laser power was kept at 0.5 mW. The system was connected to a microscope, and the laser light was coupled through an objective lens of 20×, which was used for exciting the sample as well as collecting the Raman signals. Prior to each Raman experiment calibration of the instrument was done with the Raman signal from a silicon standard centered at 520 cm$^{-1}$. Subtraction of the baseline using cubic spline interpolation was performed in order to eliminate unwanted background noise and to facilitate data analysis. SERS substrates were incubated for 2.5 h in different analyte solution at 25 ℃. Substrates were taken out and fixed onto the glass slide. SERS measurements were taken from at least eight random locations that are more than 3 mm apart with an accumulation time of 30 s.

**Apparatus and characterization.**

The Raman spectroscopy of graphene was performed using a Raman spectrometer (Horiba HR-800) with laser excitation at 532 nm (2.33 eV). Surface morphologies of G/CuNPs, TG/CuNPs and CuNPs were observed using a SEM (Hitachi S-570). The HRTEM images of G/CuNPs were obtained using a transmission electron microscopy system (JEOL, JEM-2100) operated at 100 kV. The surface compositions of G/CuNPs, TG/CuNPs and CuNPs were characterized by XRD (Rigaku D/MAX-RB). XPS was carried out on a VGESCA Lab-250 using Al Kα x-rays as the excitation source. Curve fitting of the spectra was carried out using a Lorentzian peak shape.

**Conflict of Interest:** The authors declare no competing financial interest.

**Acknowledgments**

The authors are grateful for financial support from the Shandong Province Natural Science Foundation (ZR2014FQ032, ZR2013HL049), National Natural Science Foundation of China (11474187, 61205174)

**Supporting Information Available:** XRD, SEM, EDS, XPS after thermal oxidation treatment; the linear SERS responses versus concentration of adenosine from TG/CuNPs and CuNPs; experimental procedure for fabrication of G/CuNPs and CuNPs; the analysis of effect of $H_2/CH_4$ ratio used for graphene growth. This material is available free of charge via the Internet at http://pubs.acs.org.

**Figures and Captions**

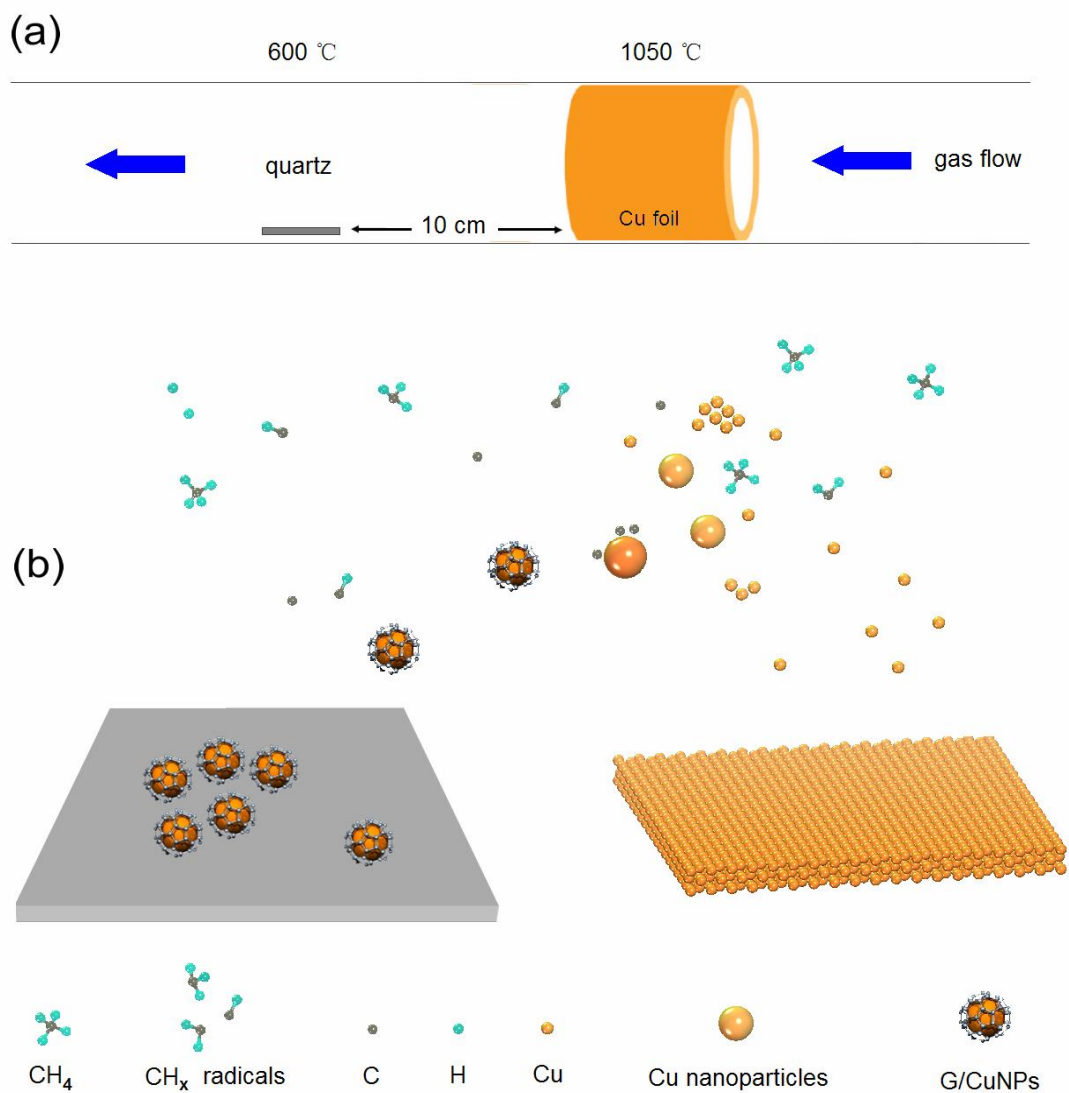

Figure 1. (a) Growth setup. A strip of Cu foil surrounding along the tube wall was used to supply Cu nanoparticles as a floating catalyst. The substrate was placed 10 cm away from the Cu strip. (b) Schematic illustration of graphene growth mechanism involving decomposition of $CH_4$ by floating Cu and H. Cu atoms are subliming from the Cu foil at 1050 °C and evolve into Cu nanoparticles at a certain concentration. Graphene starts growing on Cu nanoparticles in the mixture of $H_2$ and $CH_4$.

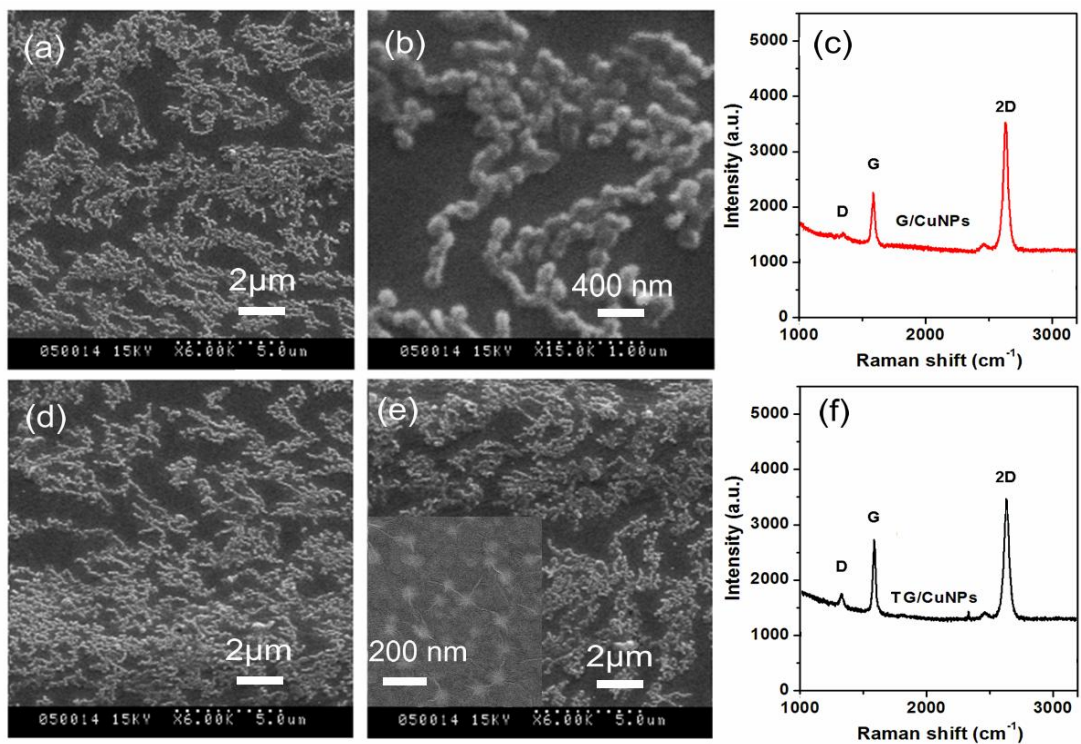

Figure 2. SEM images of G/CuNPs, CuNPs and TG/CuNPs on quartz substrate. (a) SEM image of G/CuNPs at a low magnification. (b) SEM image of G/CuNPs at a higher magnification. (c) Raman spectrum of G/CuNPs. (d) SEM image of CuNPs grown on quartz substrate. (e) SEM image of TG/CuNPs on quartz substrate. (f) Raman spectrum of CuNPs.

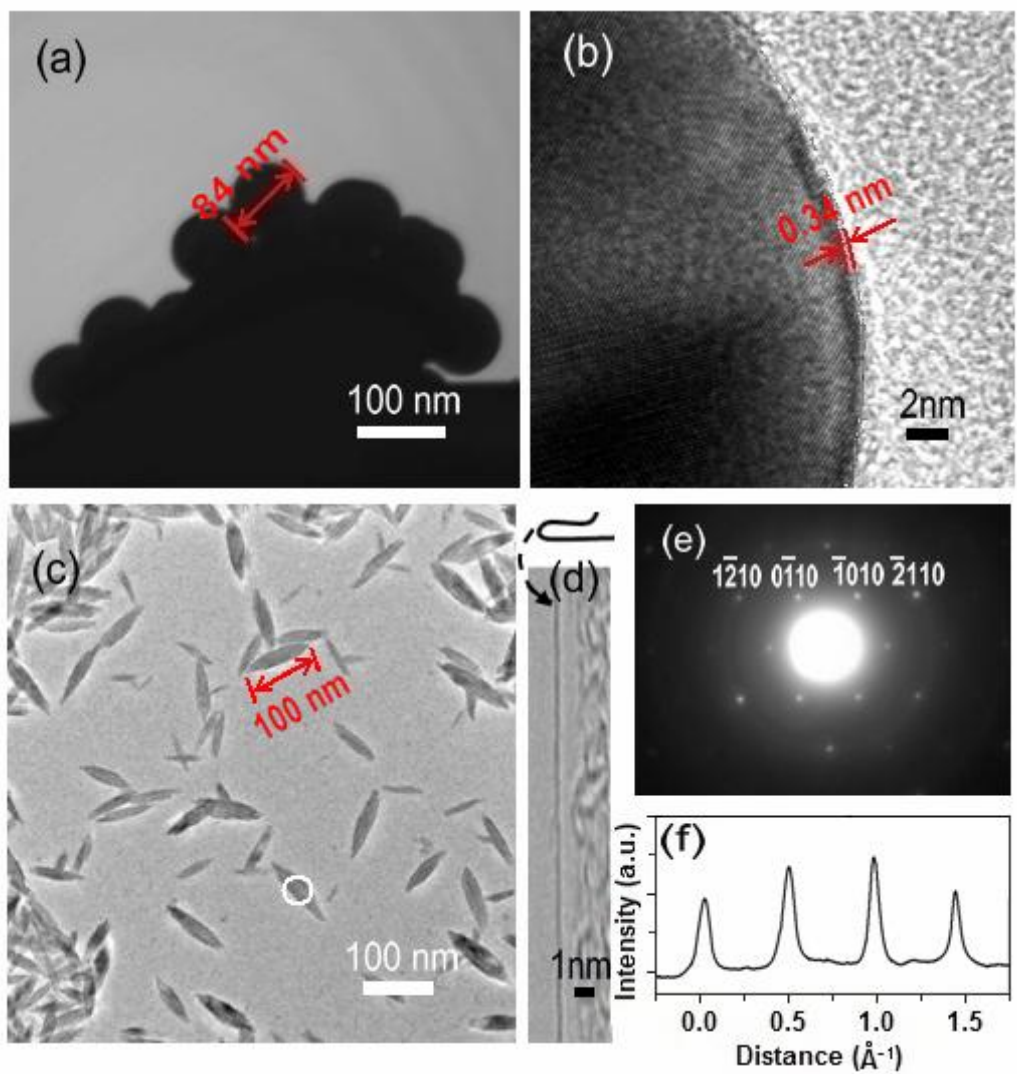

Figure 3. HRTEM images of G/CuNPs. (a) Bright-field TEM image of G/CuNPs. (b) HRTEM image of cross section of a graphene film grown on Cu nanoparticals. (c) Bright-field TEM of the graphene shell after removing the Cu nanoparticals. (d) TEM images of folded edges for monolayer graphene. (e) Electron diffraction patterns taken from the positions of the graphene sheet marked by white spots in (c). (f) Diffracted intensity taken along the $\bar{1}2\bar{1}0$ to $\bar{2}1\bar{1}0$ axis on the patterns shown in (e).

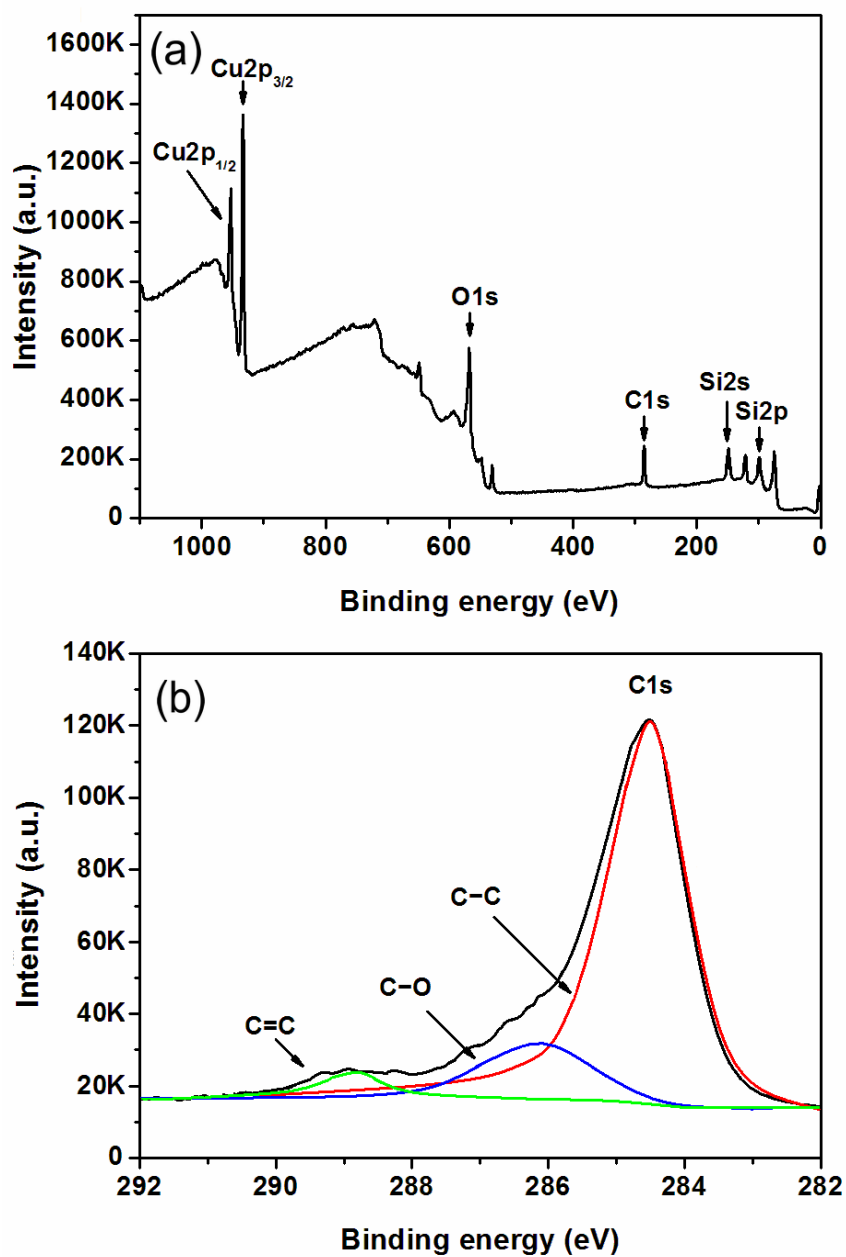

Figure 4. XPS spectra of G/CuNPs on quartz substrates. (a) Full XPS spectrum of G/CuNPs on quartz substrates. (b) Detailed C1s core-level spectra of graphene shell on Cu nanoparticles. Curve fitting of the spectra was carried out using a Gaussian-Lorentzian peak shape.

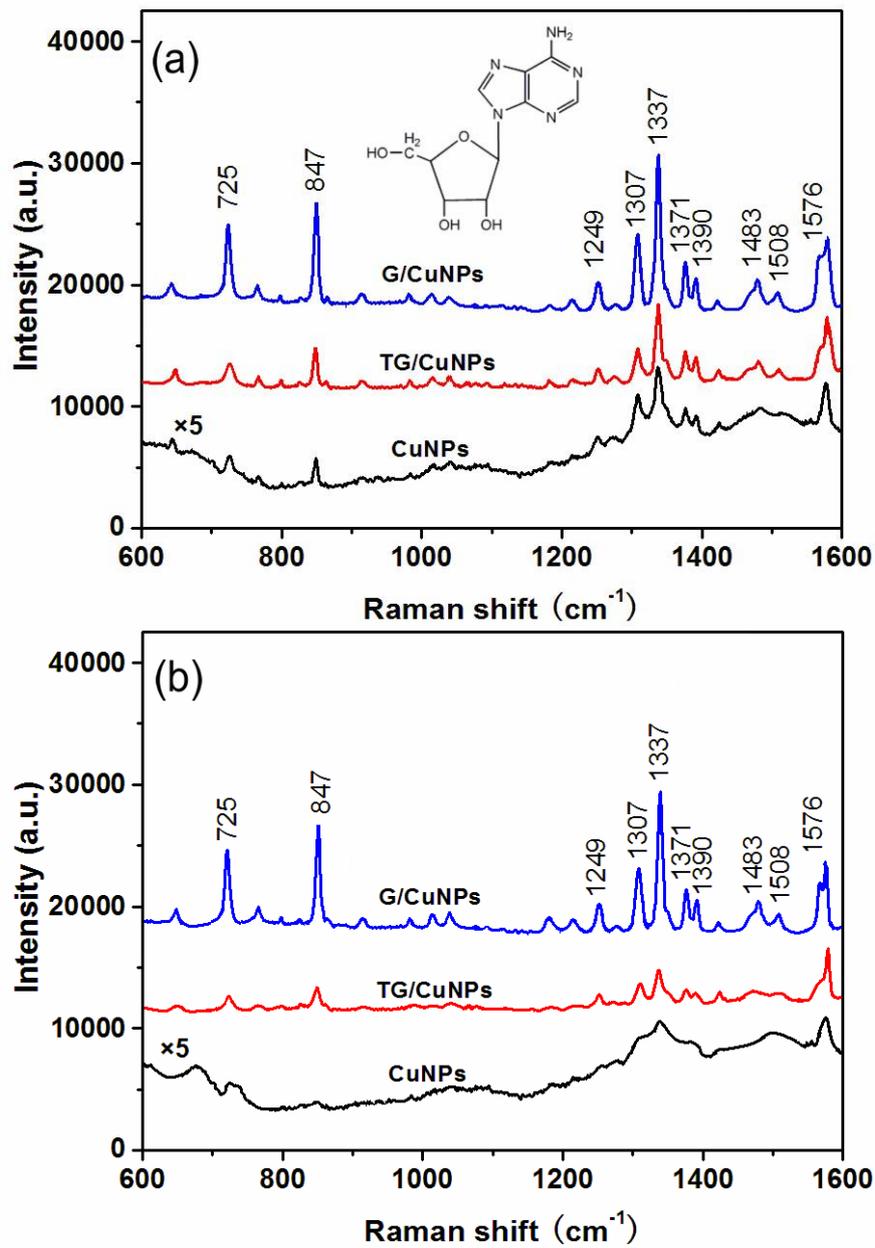

Figure 5. (a) Comparison of SERS spectra of adenosine on G/CuNPs, TG/CuNPs and CuNPs. Inset: the structural formula of adenosine molecule. (b) Comparison of SERS spectra of adenosine on G/CuNPs, TG/CuNPs and CuNPs after oxidation treatment.

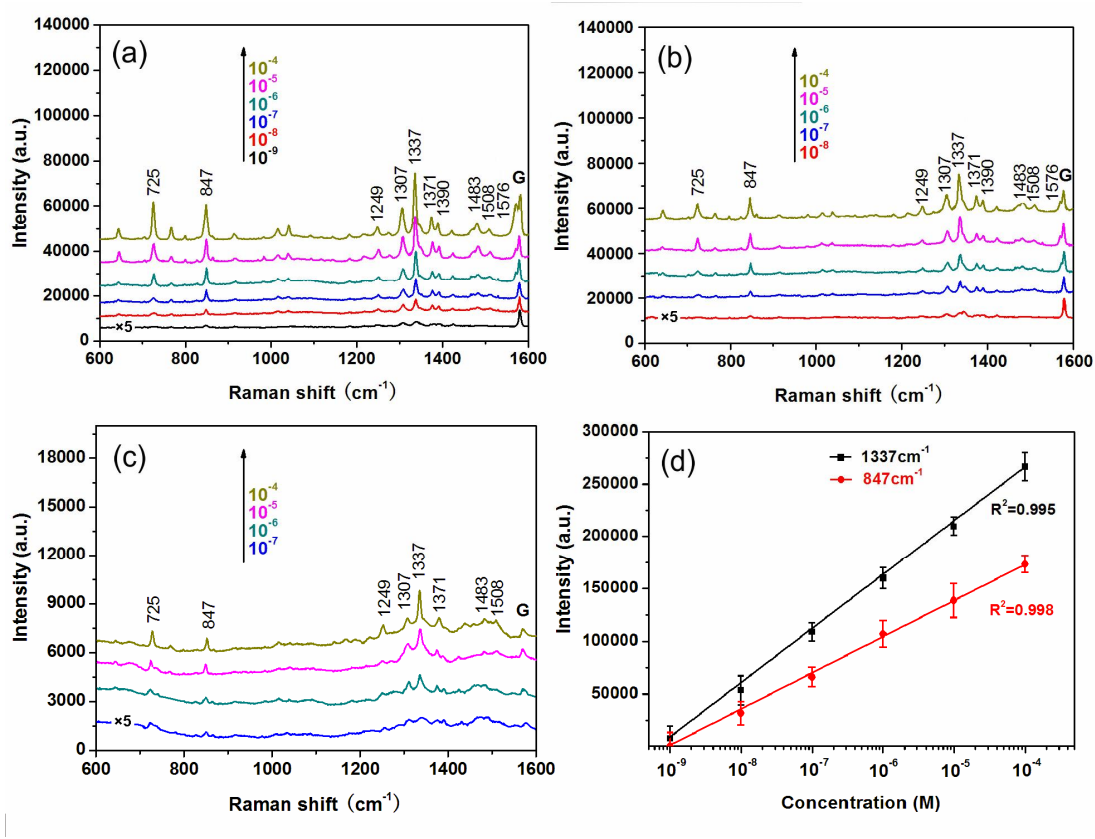

Figure 6. (a) Raman spectra of adenosine on G/CuNPs with different concentrations from $10^{-9}$ to $10^{-4}$ M. (b) Raman spectra of adenosine on TG/CuNPs with different concentrations from $10^{-8}$ to $10^{-4}$ M. (c) Raman spectra of adenosine on CuNPs with different concentrations from $10^{-7}$ to $10^{-4}$ M. (d) Raman intensity of adenosine peaks at 847 and 1337 cm$^{-1}$ on G/CuNPs, as a function of the adenosine molecular concentration with the high coefficient of determination ($R^2$) of 0.995 and 0.998, respectively.

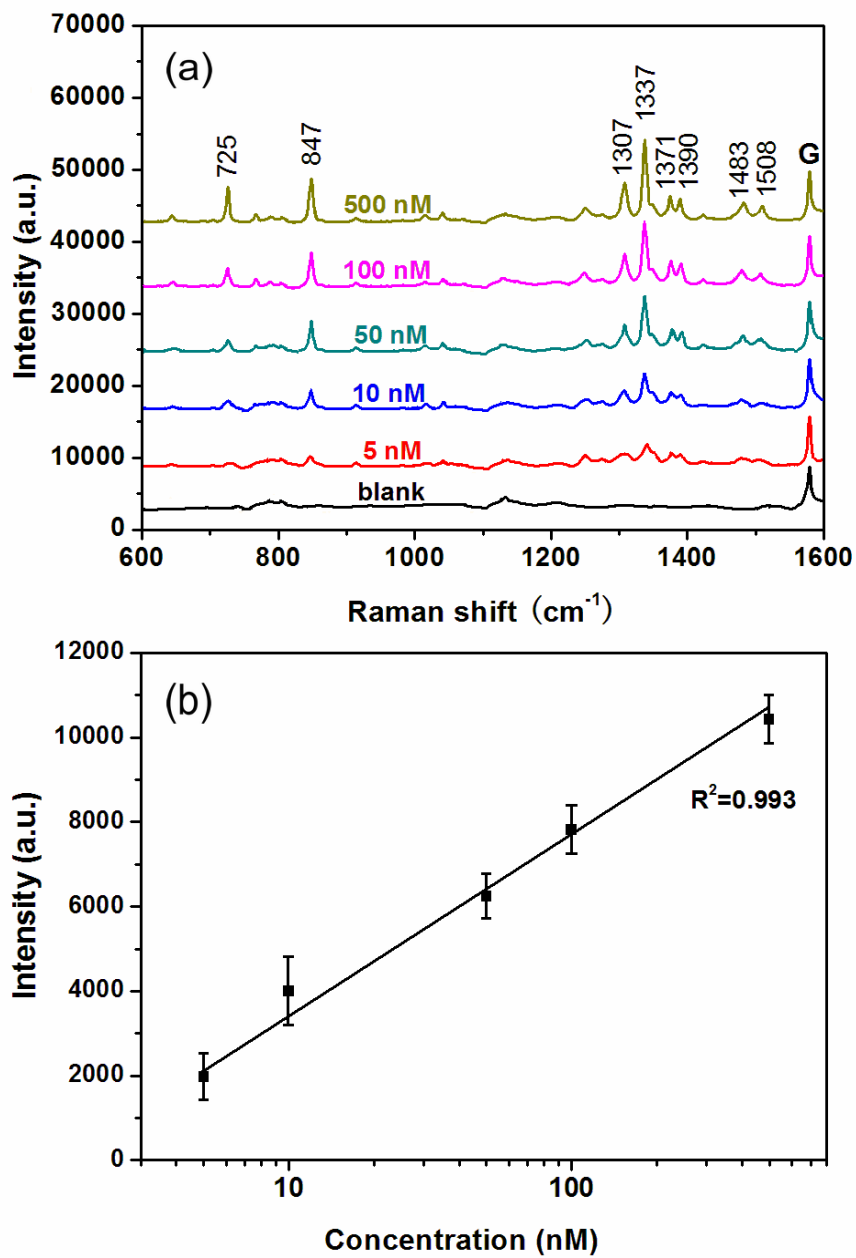

Figure 7. (a) Raman spectra of adenosine in diluted serum tested on G/CuNPs with different concentrations from 5 to 500 nM. (b) Raman intensity of adenosine in diluted serum at 1337 cm$^{-1}$ on G/CuNPs, as a function of the adenosine concentration.

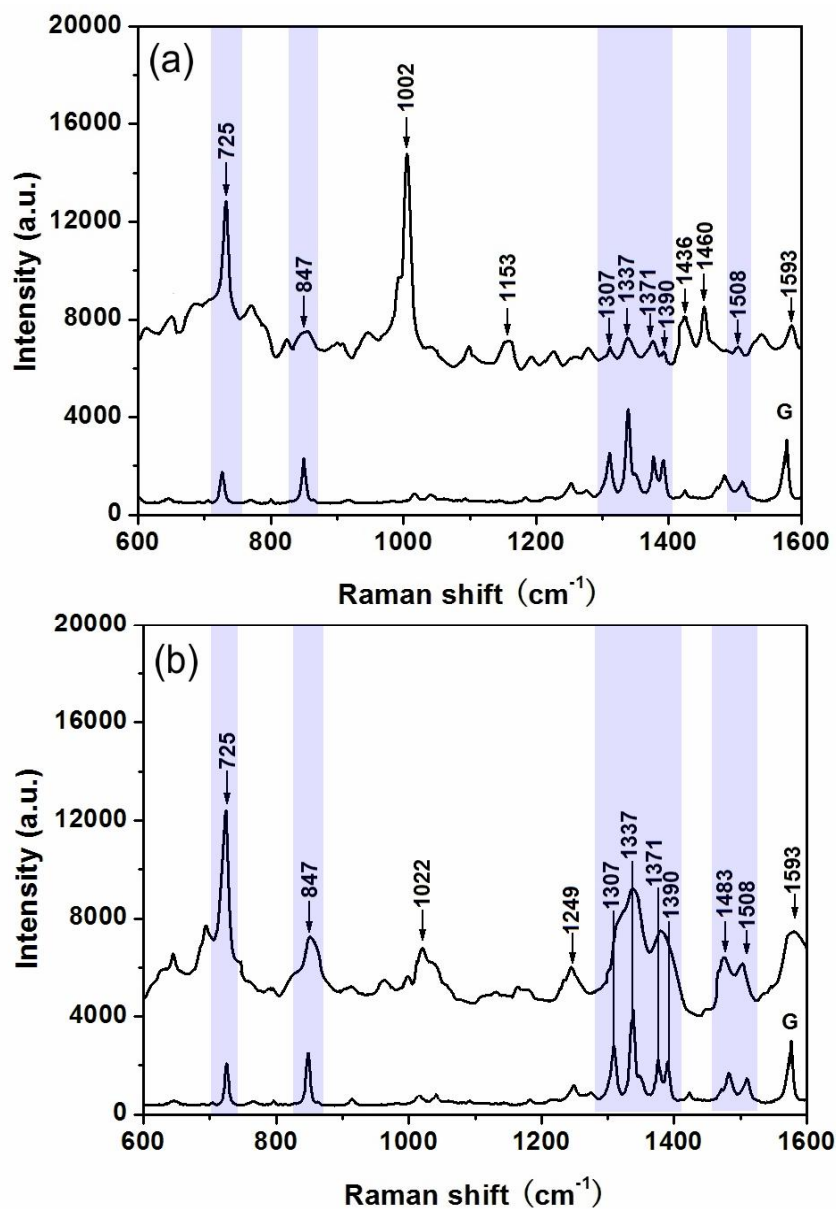

Figure 8. (a) Raman spectrum of normal human urine samples tested on G/CuNPs (top curve) and the Raman spectrum of adenosine with the concentration of 100 nM used as a reference (bottom curve). (b) Raman spectra of nucleosides excreted in normal human urine sample tested on G/CuNPs (top curve) and the Raman spectra of adenosine with the concentration of 100 nM used as a reference (bottom curve).

Supporting Information for

**Direct Growth Graphene on Cu Nanoparticles by Chemical Vapor Deposition as Surface-Enhanced Raman Scattering Substrate for Label-Free Detection of Adenosine**

Shicai Xu[1*], Baoyuan Man[2], Shouzhen Jiang[2], Jihua Wang[1], Jie Wei[3], Shida Xu[4], Hanping Liu[1], Shoubao Gao[2], Huilan Liu[1], Zhenhua Li[1], Hongsheng Li[5], Hengwei Qiu[2]

[1] College of Physics and Electronic Information, Shandong Provincial Key Laboratory of Functional Macromolecular Biophysics, Institute of Biophysics, Dezhou University, Dezhou 253023, China

[2] College of Physics and Electronics, Shandong Normal University, Jinan 250014, China

[3] Department of Neurology, Dezhou People's Hospital, Dezhou 253014, China

[4] Department of Internal Medicine, Dezhou People's Hospital, Dezhou 253014, China

[5] Department of Radiation Oncology, Key Laboratory of Radiation Oncology of Shandong Province, Shandong Cancer Hospital and Institute, Jinan 250117, China

[*] E-mail address: xushicai001@163.com (S. C. Xu)

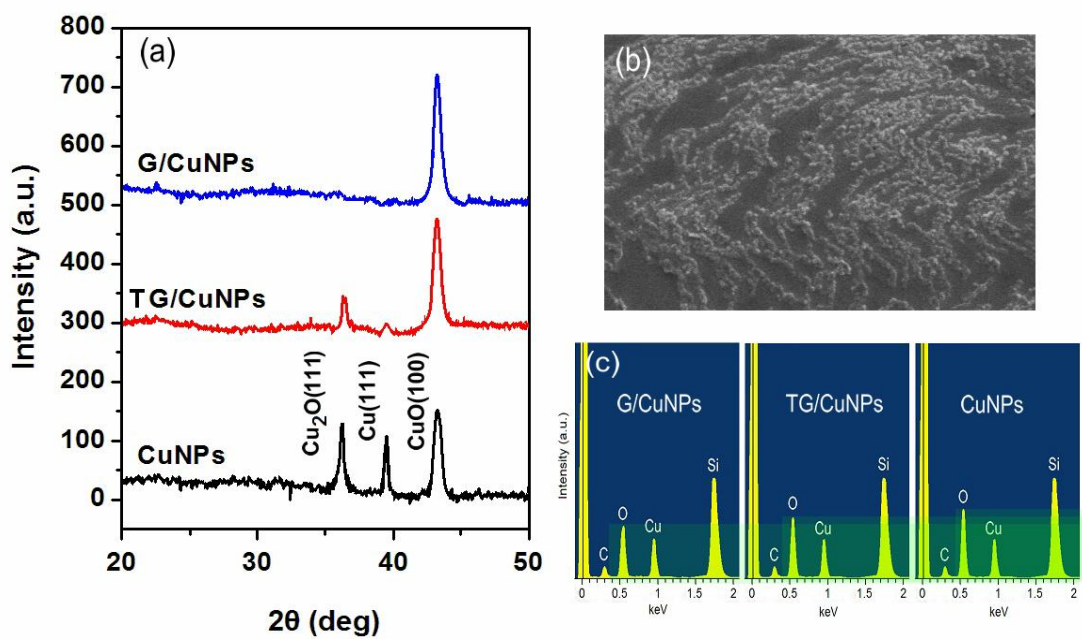

Figure S1. Stability test of the SERS substrate based on G/CuNPs, TG/CuNPs and CuNPs. (a) XRD patterns of G/CuNPs, TG/CuNPs and CuNPs after thermal oxidation treatment. (b) SEM image of CuNPs after thermal oxidation treatment. (c) EDS of G/CuNPs, G/CuNPs and CuNPs after thermal oxidation treatment.

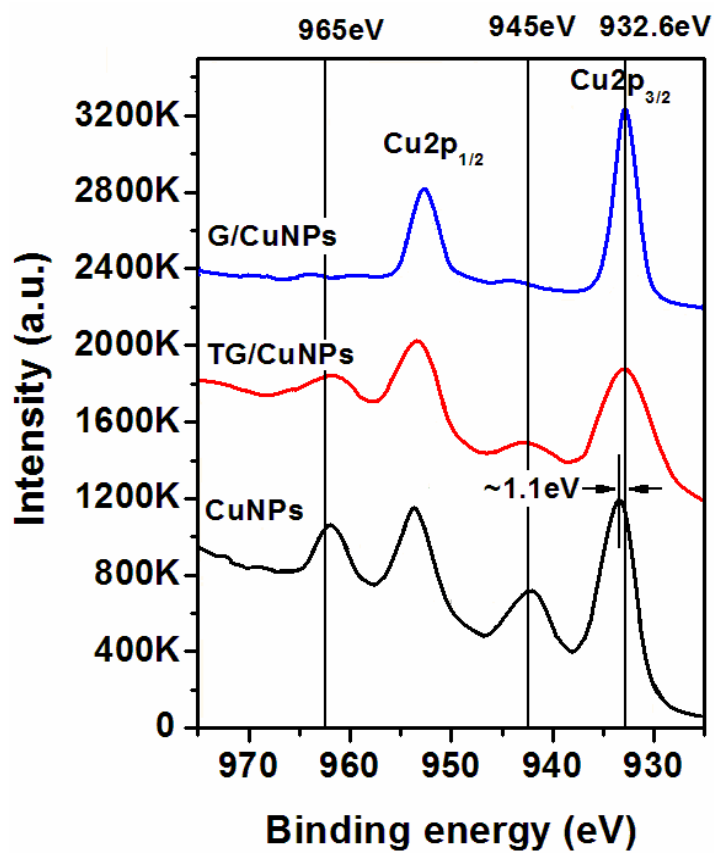

Figure S2. Spectra of Cu 2p core levels of G/CuNPs, G/CuNPs and CuNPs after thermal oxidation treatment. For TG/CuNPs and CuNPs, Shake-up features at ~945 and ~965 eV for the Cu $2p_{3/2}$ and $2p_{1/2}$ core levels are evident and diagnostic of an open $3d^9$ shell of Cu (+2).

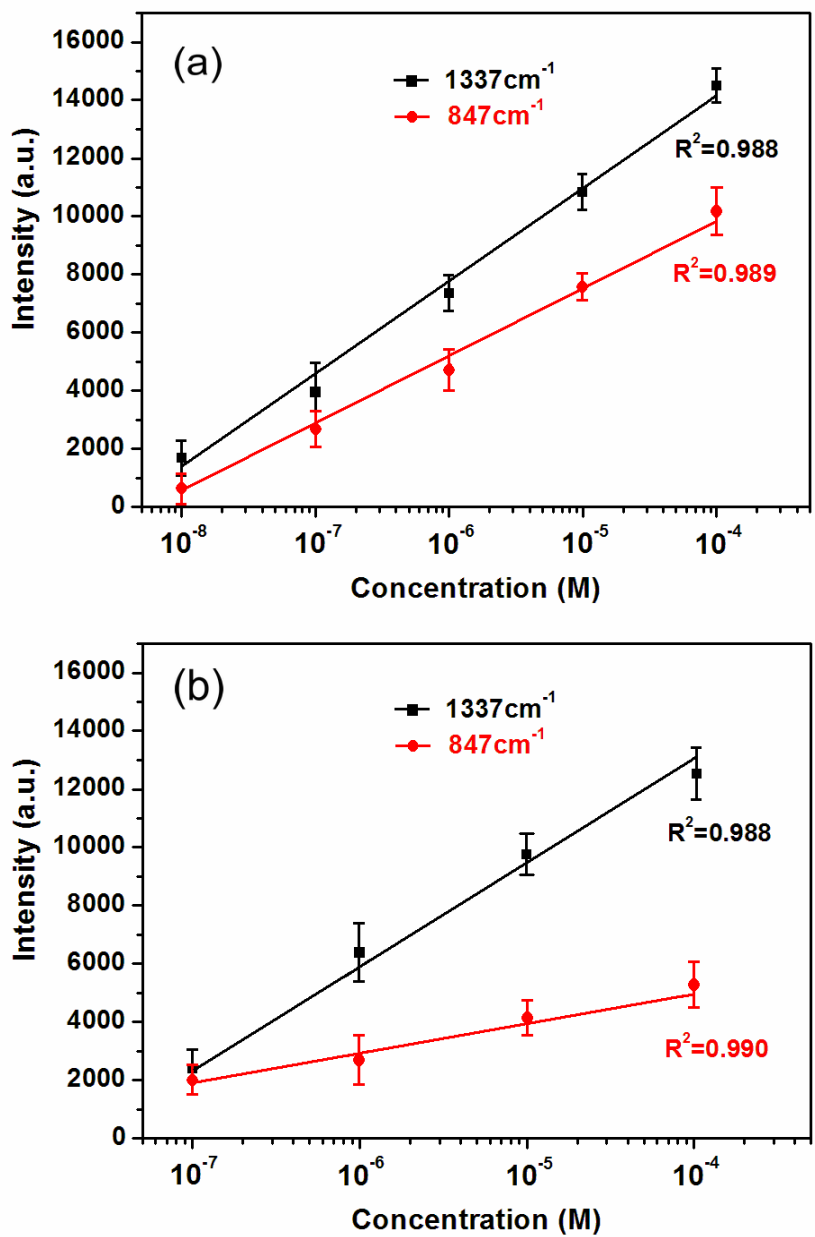

Figure S3. The linear SERS responses versus concentration of adenosine from TG/CuNPs and CuNPs. (a) Raman intensity of adenosine peaks at 847 and 1337 cm$^{-1}$ on TG/CuNPs as a function of the adenosine concentration with the coefficient of determination ($R^2$) of 0.988 and 0.989, respectively. (b) Raman intensity of adenosine peaks at 847 and 1337 cm$^{-1}$ on CuNPs as a function of the adenosine concentration with the coefficient of determination ($R^2$) of 0.988 and 0.990, respectively.

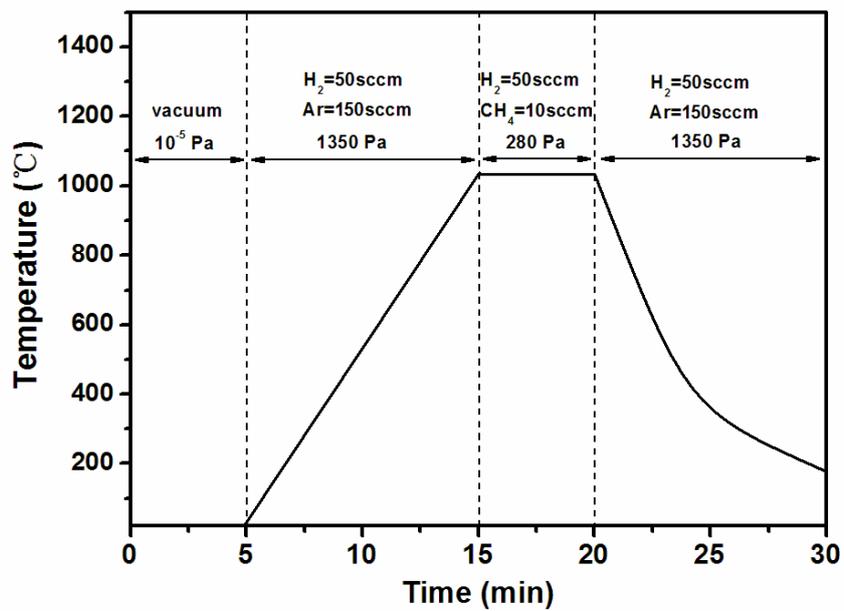

Figure S4. Experimental procedure for fabrication of G/CuNPs on flat quartz substrate.

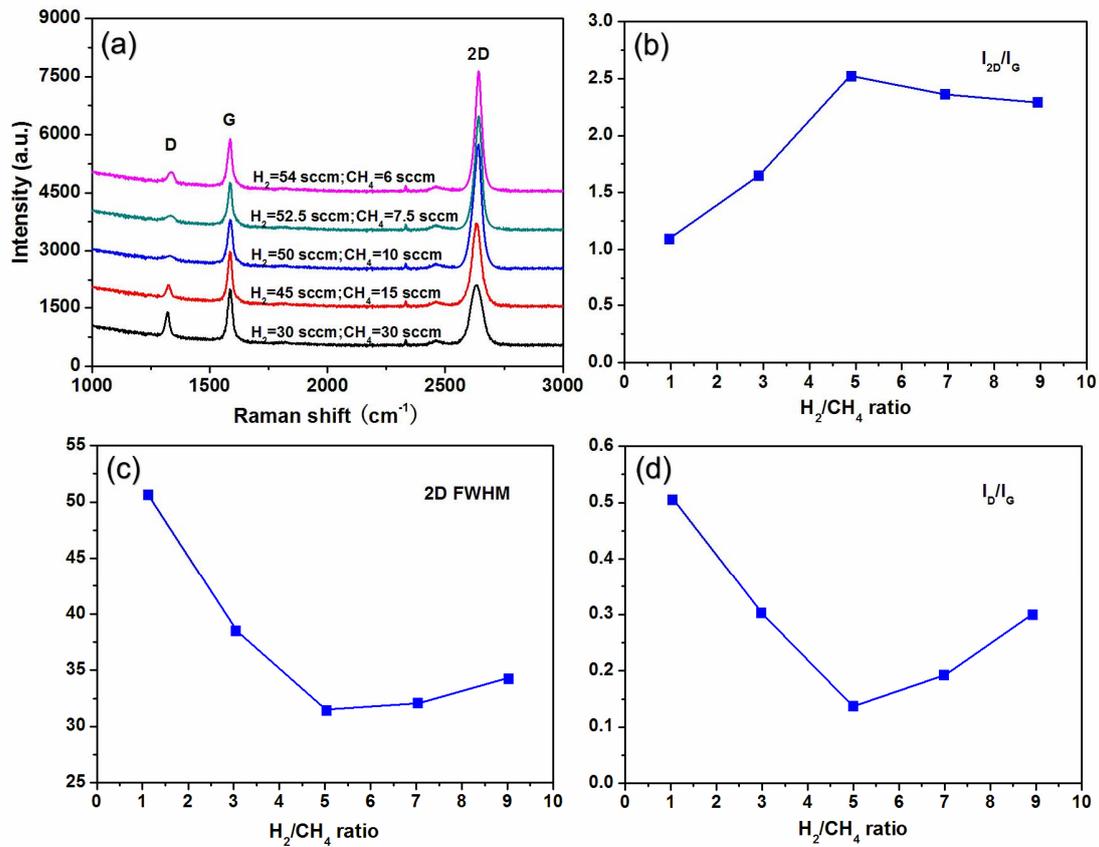

Figure S5. The effect of $H_2/CH_4$ ratio used for graphene growth. (a) Raman spectra of G/CuNPs with different flow rate of $H_2$ and $CH_4$. (b) Intensity ratio of $I_{2D}/I_G$, vs. $H_2/CH_4$ ratio. (c) Intensity ratio of $I_D/I_G$, vs. $H_2/CH_4$ ratio. (d) FWHM (the full width half maximum) of 2D peak vs. $H_2/CH_4$ ratio. The total flow rate remained 60 sccm during growth. The value of $I_{2D}/I_G$, $I_D/I_G$ and FWHM 2D vary with $H_2/CH_4$ ratio. By using 50 sccm $H_2$ and 10 sccm $CH_4$, $I_{2D}/I_G$ reaches about 2.5 and the FWHM 2D decrease to about 32, indicative of features of graphene monolayer. Also, at this $H_2/CH_4$ ratio, $I_D/I_G$ reaches the minimum value of about 0.14, indicating very low level of defect.

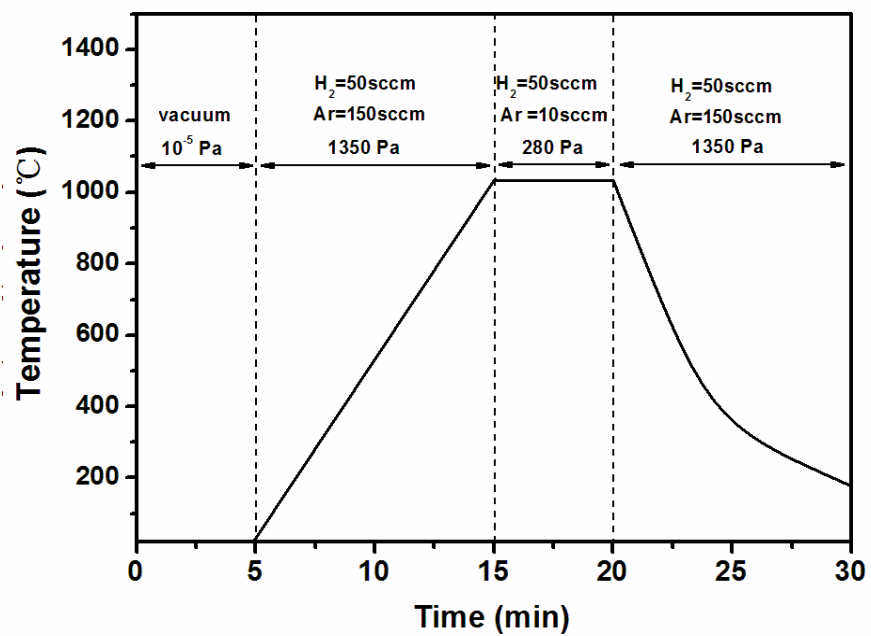

Figure S6. Experimental procedure for fabrication of CuNPs on flat quartz substrate.

.